\documentclass[12pt,letterpaper]{article}
\usepackage{epsfig}
\usepackage{setspace} 
\usepackage{footmisc}
\usepackage{graphicx}
\usepackage{amsfonts}
\usepackage{amsmath}
\usepackage{amsthm}
\usepackage{float}
\usepackage{hyperref}
\usepackage{color}
\usepackage{natbib}
\usepackage{verbatim}
\usepackage{subfig}
\usepackage{lineno}
\usepackage{soul}
\usepackage
[
	papersize={8.5in,11in},
	hmargin=1.25in,
        top=1.25in,
        bottom=1.25in
]
{geometry}

\RequirePackage{bm}

\newcommand{\footremember}[2]{%
   \footnote{#2}
    \newcounter{#1}
    \setcounter{#1}{\value{footnote}}%
}
 
\title{Modeling short-ranged dependence in block extrema with application to polar temperature data}
\author{%
    Brook T. Russell\footremember{clemson}{Clemson University School of Mathematical and Statistical Sciences, Clemson, SC 29634 (email: brookr@clemson.edu)}%
    \and Whitney K. Huang
 }

\begin{document}
\maketitle

\begin{abstract}
The block maxima approach is an important method in univariate extreme value analysis. While assuming that block maxima are independent results in straightforward analysis, the resulting inferences maybe invalid when a series of block maxima exhibits dependence. We propose a model, based on a first-order Markov assumption, that incorporates dependence between successive block maxima through the use of a bivariate logistic dependence structure while maintaining generalized extreme value (GEV) marginal distributions. Modeling dependence in this manner allows us to better estimate extreme quantiles when block maxima exhibit short-ranged dependence. We demonstrate via a simulation study that our first-order Markov GEV model performs well when successive block maxima are dependent, while still being reasonably robust when maxima are independent. We apply our method to two polar annual minimum air temperature data sets that exhibit short-ranged dependence structures, and find that the proposed model yields modified estimates of high quantiles.


\end{abstract}



\section{Introduction} \label{sect:Intro}

In recent years, there has been an increased focus on modeling extremes of environmental variables. \textit{Extreme value theory} (EVT)
provides a probabilistic framework for performing statistical inference on the far upper tail of distributions, and is therefore useful in a wide variety of environmental applications. Examples include  modeling extreme temperatures \citep{huang2016,stein2020a,stein2020b,o2020}, precipitation extremes \citep{huang2019,russell2020,hazra2020,fix2020}, and extremes in hydrology \citep{towe2019,beck2020}.

In the analysis of univariate extremes, the block maxima approach \citep{Coles01a,Gumbel1958} is among the most commonly employed methods. Under this framework of analysis, briefly outlined in Section \ref{sect:ExtremesReview}, renormalized block maxima can be shown to converge to the generalized extreme value (GEV) distribution under certain conditions. Inference is typically performed by assuming that a series of block maxima are independent GEV realizations. In many applications, this assumption of independence among block maxima is quite reasonable \citep[e.g.,][]{huang2016}, given that the block size is sufficiently large and the (within block) serial dependence is relatively weak; however, there are cases where dependence between consecutive block maxima is exhibited \citep[e.g.,][]{zhu2019}. In these instances, traditional block maxima analysis will ignore this dependence, potentially disregarding important information and leading to invalid  inference.

In this work, we are motivated by series of block minima that appear to exhibit short-ranged \textit{asymptotic (tail) dependence}. Informally, two variables are asymptotically dependent if the probability of the event that both are at their most extreme levels simultaneously is non-zero. Our motivating data sets consist of annual minimum temperatures at Arctic and Antarctic research stations, and both exhibit the aforementioned short-ranged asymptotic dependence structure. That is, consecutive block minima appear to show dependence, but this (estimated) dependence diminishes beyond a lag of one (e.g., more than one year apart. See Fig.~\ref{fig:FaradayPlot}). Although we are interested in modeling block \textit{minima}, we note that we begin by developing methodology for block \textit{maxima} in this manuscript. This approach is justified by the fact that methods for analysis of block maxima are easily applied to block minima after the response variable is negated.

\cite{zhu2019} analyze the same Antarctic data set that we consider here and conclude that the series of annual minima exhibits short-ranged dependence, primarily at lag one. However, their approach is based on using a Gaussian copula \citep{joe1997,nelsen2006} to model dependence. Although copula-based methods are flexible in terms of modeling dependence in general, Gaussian copula based models are incapable of modeling asymptotic dependence \citep{sibuya1959}. 
In extreme value analysis, asymptotic dependence is often the type of dependence that is of primary interest. We give a brief overview of asymptotic dependence in Section \ref{sect:ExtremesReview}; \cite{Coles01a} and \cite{Resnick2007}  provide additional details for the interested reader.

In order to model such asymptotic dependence among block maxima, one potential approach is to employ multivariate extreme value methods \citep[see, for example, Ch.\ 8 of][]{Coles01a}. Unfortunately, characterizing dependence for extremes is nontrivial even for moderate dimension (e.g., dimension $d \geq 3$), and model parameters may be difficult to estimate and interpret. Another option is to utilize a max-stable process based approach \citep{smith1990,zhang2004,davison2012}. Regrettably, the corresponding likelihood functions are not easily expressed because of a combinatorial explosion of terms as a function of the dimension $d$ and therefore the resulting model is challenging to fit \citep{castruccio2016}. This complication with the likelihood function makes both maximum likelihood and Bayesian inference approaches impractical.

In this work we propose a modeling procedure specifically for block maxima with a short-ranged asymptotic dependence structure. Our approach is based on a first-order Markov assumption, and offers several attractive properties: GEV marginal distributions, a likelihood function with an easily expressed closed form that makes both frequentist and Bayesian inference relatively straightforward, and short-ranged \textit{asymptotic dependence} of block maxima.

This manuscript is organized in the following manner. In Section \ref{sect:ExtremesReview} we give a brief background in univariate and bivariate extremes used in this work. We describe our method for modeling dependent block maxima in Section \ref{sect:DepGEV} and give the results of a simulation study in Section \ref{sect:SimStudy}. Our analysis of temperature data is given in Section \ref{sect:Analysis}. We conclude with a discussion in Section \ref{sect:Discussion}.

\section{Review of Univariate and Bivariate Extremes} \label{sect:ExtremesReview}
In this section, we provide background for univariate and bivariate extreme value analysis. We introduce the block maxima approach in Section \ref{sec2.1}, as our method of analysis is built within this framework. An asymptotic dependence measure and the bivariate logistic dependence model is described in Section \ref{sec2.2}. 

\subsection{Univariate Extremes and the Block Maxima Approach} \label{sec2.1}
\cite{fisher1928} and \cite{gnedenko1943} provide the theoretical basis for modeling block maxima, the maxima taken from sequences of $n$ independent and identically distributed random variables $(X_1,\ldots,X_n)$, with ``block length'' $n$ sufficiently large, through the use of an extreme value distribution. Specifically, let $M_n = \max \{X_1, \ldots, X_n\}$ denote the block maxima. If there exist sequences $\{a_n > 0\}$ and $\{b_n\}$ such that


\begin{equation}
P\left( \frac{M_n - b_n}{a_n} \leq z \right) \stackrel{n \to \infty}{\longrightarrow} G(z)
\end{equation}
for a non-degenerate distribution  $G$, then $G$ must belong to the (reversed) Weibull family, the Gumbel family, or the Fr\'echet family. The GEV distribution is a three parameter distribution that includes these three families as special cases. We let $Z$ denote the (non-degenarate) asymptotic distribution of the block maximum (e.g., annual maximum) of a variable of interest (e.g., daily temperature with block size $n = 365$). For $Z \sim \text{GEV}(\mu,\sigma,\xi)$, 
its distribution function is defined such that 
\begin{equation} \label{eq:DefGEV}
P(Z < z) = \exp \left( - \left(1 + \xi \left( \frac{z-\mu}{\sigma} \right) \right)_+^{-1/\xi} \right),
\end{equation}
where $c_+ := \max \{ c,0 \}$. 
The GEV parameters are referred to as the \textit{location} parameter $\mu \in \mathbb{R}$, the \textit{scale} parameter $\sigma>0$, and the \textit{shape} parameter $\xi \in \mathbb{R}$. All three parameters are involved in modeling extremes, but the shape is especially important as it determines the nature of the tail. If $\xi < 0$, the tail will be bounded and the GEV becomes the Weibull. If $\xi > 0$, the tail will be heavy and the GEV becomes the Fr\'echet. If $\xi \rightarrow 0$, the resulting  tail will be light and the GEV becomes the Gumbel. \cite{Coles01a} offers an introductory resource for the analysis of univariate extremes. We note that \cite{leadbetter1983} show that the independence assumption on $\left\{X_{1}, \ldots, X_{n}\right\}$ can be relaxed for weakly dependent stationary time series. \cite{einmahl2016} extend the theory to non-identically distributed observations
when distributions of $\left\{X_{1}, \ldots, X_{n}\right\}$ share a common absolute maximum.


In practice, block maxima are extracted where the blocks are produced by dividing the data record (e.g., a time series of certain climate variable) into non-overlapping periods. If the blocks are thought to be large enough, the series of block maxima may be considered GEV realizations and can be used to estimate the corresponding GEV parameters. This can be done using a likelihood \citep{prescott1980} or a moment based method \citep{hosking1985}, or alternatively via a Bayesian approach \citep{coles1996}. For the case where analysis of minima is of interest, researchers can use this type of approach after negating the series of block maxima.

In application, researchers are often interested in estimating the $(1-p)^{th}$ quantile of $Z$, the distribution of block maxima, with a ``small'' $p$, say 0.05 or 0.01.
For $Z \sim \text{GEV}(\mu,\sigma,\xi)$, where $\mu$, $\sigma$, and $\xi$ are known, this quantile is given by
\begin{equation} \label{eq:GEVquantile}
Z_p(\mu,\sigma,\xi) =
\begin{cases}
\mu - \frac{\sigma}{\xi} (1 - \{- \log (1-p)  \}^{-\xi} ) \text{ for } \xi \neq 0 \\
\mu - \sigma \log \{- \log (1-p)   \} \text{   for } \xi = 0.
\end{cases}
\end{equation}
After estimating the three GEV parameters based on sample data, the $(1-p)^{th}$ quantile can be estimated via the plug-in estimate $\widehat{Z}_p = Z_p(\hat{\mu},\hat{\sigma},\hat{\xi})$. Importantly, when the data are composed of annual maxima, one can  estimate the $r$-year return level that is associated with exceedance probability $p$, and is denoted  
$\widehat{RL}_{r} = \widehat{Z}_p$. Here, the return period is $r=1/p$ years. For example, the exceedance probability of .02 corresponds to a return period of $1/.02=50$ years. 

Quantifying uncertainty in return level estimates can be performed in several ways. A delta method based approach can be used, but the resulting confidence intervals are known to perform poorly when $p$ is small \citep{Coles01a}. For likelihood based inference, a profile likelihood method is sometimes recommended. When Bayesian inference is employed, credible intervals for return levels can be produced from the Markov chain Monte Carlo (MCMC) output. 
\cite{coles1996} provide a discussion of this issue in the extreme value analysis context. 

\subsection{Bivariate Extremes and Asymptotic Dependence} \label{sec2.2}
When performing an extreme value analysis of bivariate random vectors, describing asymptotic dependence is often of primary interest. In this section, we briefly introduce asymptotic dependence and discuss a parametric model for asymptotic dependence.

\subsubsection{Characterizing Asymptotic Dependence}
For random variables $X$ and $Y$ with their corresponding cumulative
distribution function $F_{X}$ and $F_{Y}$, define the \textit{tail dependence coefficient} $\chi$ \citep{coles1999}
\begin{equation} \label{eq:ChiDef}
\chi = \lim_{u \rightarrow 1^{-}} P(F_{Y}(Y) > u | F_{X}(X)>u).
\end{equation}
If $\chi > 0$, the two variables are termed \textit{asymptotically dependent}; the asymptotic independence case is implied when $\chi = 0$. Determining the presence and the strength of asymptotic dependence is important in many applications. For example, experiencing extreme levels of storm surge and precipitation simultaneously may result in much greater damage compared to one of these variables reaching extreme levels by its self.

Typical association metrics, such as Pearson's correlation coefficient, may be useful for describing association in the bulk of the data; however, they often perform poorly in terms of describing asymptotic dependence. It is also important to note that bivariate Gaussian random variables with a correlation coefficient less than one are asymptotically independent \citep{sibuya1959}. Therefore, one should carefully consider whether to use Gaussian copulas to model random variables that may exhibit asymptotic dependence.

\subsubsection{A Block Maxima Approach for Analysis of Bivariate Extremes}
Assume that $\{ (X_i,Y_i) \}_{i \in \mathbb{N}}$ is a sequence of independent bivariate random vectors with joint distribution function $F_{X,Y}(x,y)$. Define the vector of componentwise maxima, $\bm M_n = (M_{x,n},M_{y,n})$, where $M_{x,n} = \underset{i \in \{1,\ldots,n\}}{\max} \{X_i\}$ and $M_{y,n} = \underset{j \in \{1,\ldots,n\}}{\max} \{Y_j\}$.
Importantly, we note that the index for which the maximum of the $X_i$s occurs is not necessarily the same index for which the maximum of the $Y_j$s occurs. That is, 
$\underset{i \in \{1,\ldots,n\}}{\text{argmax}} ~ X_i$ is not necessarily the same as $\underset{j \in \{1,\ldots,n\}}{\text{argmax}} ~ Y_j$, and therefore $\bm M_n$ may not appear in the original data set. As is standard in extreme value analysis, we transform the marginal distributions of $M_{x,n}$ and $M_{y,n}$ such that both have the unit Fr\`echet distribution (a special case of $\text{GEV}(\mu = 1, \sigma = 1, \xi=1)$), with distribution function given by 
\begin{equation}
 P(Z \leq z) = \exp(-z^{-1}) \text{ for } z>0.   
\end{equation}
The use of such marginal transformations is theoretically justified \citep{Resnick2007} and allows for describing asymptotic dependence in a more straightforward manner.

If $P(M_{x,n} \leq x, M_{y,n} \leq y) \stackrel{d}{\rightarrow} G(x,y)$ for non-degenerate $G$, then $G$ will have the form
\begin{equation}
 G(x,y) = \exp \left( - V(x,y) \right)   
\end{equation}
for $x,y > 0$. The function $V$ can be expressed in the form  
\begin{equation}
V(x,y) = 2 \int_0^1 \max \left( \frac{w}{x} , \frac{1-w}{y} \right) dH(w),
\end{equation}
where $H$ is a non-negative measure that determines the dependence and satisfies 
\begin{equation}
\int_0^{1} w\, dH(w) = \int_{0}^{1} (1-w)\, dH(w) = 1.
\end{equation}
\citep{tawn1988,coles1991,coles1994,ledford1997}.

One approach is to model $H$ using $H_{\bm \theta}$, a parametric family with parameters $\bm \theta \in \bm \Theta$.
In this work, we utilize the logistic (as known as Gumbel) family \citep{tawn1988} where the dependence structure is determined by a single parameter $0 \leq \alpha \leq 1$. The joint cumulative distribution function is given by 
\begin{equation}
G_{\alpha}(x,y) = \exp \{ - (x^{-1/\alpha} + y^{-1/\alpha})^\alpha \}.
\end{equation}
Under this parametric modeling assumption, the parameter $\alpha$ determines the nature of the asymptotic dependence between the corresponding maxima, and a smaller parameter value implies a higher degree of tail dependence. As $\alpha \rightarrow 1$, the two maxima become asymptotically independent; as $\alpha \rightarrow 0$, they exhibit perfect dependence. We also note that under the logistic modeling assumption, there is a deterministic relationship between $\alpha$ and the parameter $\chi$ from (\ref{eq:ChiDef}), given by $\chi = 2 - 2^\alpha$ \cite[see][p.~16, Example 3.3]{embrechts2001}. We note that $\chi= \lim_{u \rightarrow u^*} P(Y > u | X > u) = 2-2^1 = 0$ for the asymptotically independent case, and $\chi = 2-2^0=1$ for the perfect dependence case.

\section{Modeling Dependent Block Maxima} \label{sect:DepGEV}

In this section, we outline a first-order Markov based model to account for short-ranged dependence among block maxima, and discuss how it is possible to use such a model to estimate conditional tail quantiles. \cite{smith1997} and \cite{smith1992} also suggest a Markov dependence structure, but in the context of threshold exceedances.

\subsection{Outlining the First-order Markov GEV Model} \label{sect:FirstOrderMarkov}
Let $\{Z_t\}_{t=1,\ldots,n}$ be a series of block maxima based on blocks with a large number of observations. At this point, we assume that each $Z_t$ has the unit Fr\`echet distribution, and each marginal distribution function is therefore given by

\begin{equation}
  F_{Z_t}(z_t) = \exp\{-z_t^{-1}\},  
\end{equation}
leading to the marginal density function
\begin{equation}
f_{Z_t}(z_t) = z_t^{-2} \exp\{-z_t^{-1}\}.
\end{equation}

We again emphasize that the unit Fr\`echet distribution is a special case of the GEV, defined in (\ref{eq:DefGEV}). Under our proposed modeling procedure, asymptotic dependence between consecutive block maxima is modeled by assuming that $Z_t$ and $Z_{t+1}$ have a bivariate extreme value distribution with logistic dependence structure. As we are only interested in modeling data with short-ranged dependence, we make a first-order Markov assumption; that is, we assume that $Z_t$ and $Z_{t+k}$ are conditionally independent given $Z_{t+k-1}$ for $k > 1$ and $t = 1, \ldots, n-k$. 

Under the bivariate logistic dependence structure described in Section \ref{sec2.2}, the joint distribution function between consecutive observations is given by
\begin{equation}
F_{Z_t,Z_{t+1}}(z_t,z_{t+1} | \alpha) = \exp \{-(z_t^{-1/\alpha}  +  z_{t+1}^{-1/\alpha})^{\alpha} \},
\end{equation}
and their joint density is given by
\begin{align}
f_{Z_t,Z_{t+1}}(z_t,z_{t+1} | \alpha) &=
F_{Z_t,Z_{t+1}} (z_t,z_{t+1}) (z_{t} z_{t+1})^{-1/\alpha} (z_{t}^{-1/\alpha} + z_{t+1}^{-1/\alpha})^{-2 + \alpha} \nonumber \\
            &  ~~~~~ \times (\alpha^{-1} - 1 + (z_{t}^{-1/\alpha} + z_{t+1}^{-1/\alpha})^{\alpha} ).
\end{align}
We can then obtain the likelihood function 
\begin{linenomath*}
\begin{align}
\mathcal{L}(\alpha | \bm Z) &= f_{\bm Z}(\bm z | \alpha) \nonumber \\ 
        &= f_{ Z_1}(z_1) ~ \prod_{t=1}^{n-1} f_{ Z_{t+1}}(Z_{t+1} | Z_{t}, \alpha) \nonumber \\ 
        &= f_{Z_1 }(z_1) ~ \prod_{t=1}^{n-1} \frac{f_{Z_t,Z_{t+1}}(z_t,z_{t+1} | \alpha)}{f_{Z_t}(z_t | \alpha)},
\end{align}
\end{linenomath*}
relying on the first-order Markov assumption. This implies that the log likelihood for the series is given by
\begin{align*}
\log(\mathcal{L}( \alpha | \bm Z)) &= \log(f_{Z_1}(z_1 | \alpha)) +  \sum_{t=1}^{n-1} \log \left(\frac{f_{Z_t,Z_{t+1}}(z_t,z_{t+1} | \alpha)}{f_{Z_t}(z_t | \alpha)}\right) \\
                         &= \sum_{t=1}^{n-1} \log (f_{Z_t,Z_{t+1}}(z_t,z_{t+1} | \alpha)) - \sum_{t=2}^{n-1} \log (f_{Z_t}(z_t | \alpha)).
\end{align*}
As the likelihood function has a relatively simple closed form, inference can be performed in a straightforward manner via maximum likelihood, and Bayesian inference is also possible. Importantly, inference on the dependence parameter $\alpha$ may yield valuable information regarding the degree to which consecutive block maxima exhibit dependence.

To this point, we have assumed unit Fr\'echet marginal distributions. To allow for arbitrary GEV marginal distributions in analysis, we rely on the following. Assume that $Z$ is a unit Fr\'echet random variable, and consider
$Y = \mu + \sigma (Z_+^\xi - 1)/\xi$ 
for $\mu \in \mathbb{R}$, $\sigma > 0$, and $\xi \in \mathbb{R}$. It is straightforward to show that 
\begin{equation} \label{eq:ArbitraryGEV}
Y \sim \text{GEV}(\mu,\sigma,\xi).
\end{equation}
In extremes, it is common practice to use this relation to perform transformations of the marginals in order to conduct analysis with unit Fr\'echet marginal distributions.

\subsection{Estimating Conditional Tail Quantiles}
When block maxima are assumed to be independent, estimating an upper tail quantile (or the corresponding return level) of interest is straightforward, and is simply a function of the three GEV parameter estimates; this relation is given in Equation (\ref{eq:GEVquantile}). In the case where block maxima exhibit short-ranged dependence, it may be possible to improve estimates of upper tail quantiles at time $t+1$ by incorporating information regarding the block maximum at time $t$.

Assume that the first-order Markov based model from Section \ref{sect:FirstOrderMarkov} holds. Given that $Z_t = z_t$, the distribution of $Z_{t+1}$ is given by 
\begin{equation} \label{eq:CondCDF}
    F_{Z_{t+1} | Z_t = z_t}(z_{t+1}|\alpha) = P(Z_{t+1} \leq z_{t+1} | Z_t = z_t,\alpha) = \frac{1}{f_{Z_t}(z_t)} \int_0^{z_{t+1}} f_{Z_t,Z_{t+1}}(z_t,v|\alpha) dv. 
\end{equation}
In order to obtain the desired upper tail quantile, one simply needs to find the value 
\begin{equation} \label{eq:CondTailQtile}
\inf \{ z_{t+1} > 0 | F_{Z_{t+1} | Z_t = z_t}(z_{t+1}|\alpha) \geq 1 - p \}.
\end{equation}
We are not able to find a closed form for the conditional distribution function in (\ref{eq:CondCDF}); therefore, we employ numerical methods to obtain tail quantile estimates in practice. 

Quantifying uncertainty in tail quantile estimates presents additional challenges compared to the independent block maxima case. A delta method based approach presents similar downsides as in the independent block maxima case, and profile likelihood based methods are difficult to implement. However, Bayesian inference methods yield straightforward means for producing the desired credible intervals. For this reason, we employ a Bayesian approach in this work.

\section{Simulation Study} \label{sect:SimStudy}
In order to assess the first-order Markov GEV model's ability to estimate extreme quantiles for both dependent and independent block maxima data, we undertake a simulation study. In this simulation study, we consider three data generating processes: a stationary independent GEV process, a stationary first-order Markov GEV process, and a stationary moving average process of order two, abbreviated MA(2). For all three processes, the marginal GEV location and scale parameters are taken to be 0 and 1 (respectively); in order to approximate the shape parameter in our polar temperature data application, we choose the shape parameter for both processes to be -0.1, reflecting a typical shape parameter value for near-surface air temperature extremes. For the first-order Markov GEV process, we set the dependence parameter $\alpha$ to be 0.7,  
corresponding to a moderate level of asymptotic dependence between consecutive block maxima, and similar to what we observe in our data application. 
Realizations for the first-order Markov GEV process are obtained via the use of inverse transform sampling using the conditional distribution function in Equation (\ref{eq:CondCDF}). The MA(2) process that we consider is defined by
\begin{equation} \label{eq:MA2}
X_t = W_t + 0.45 W_{t-1} + 0.075 W_{t-2},
\end{equation}
for $t \in \mathbb{N}$ and where $W$ is iid Gaussian white noise with unit variance. The resulting series is transformed to have GEV(0,1,-0.1) marginal distributions via probability integral transformations, similar to \cite{zhu2019}. We note that this MA(2) process has a true correlation of approximately 0.42 at lag one, and a true correlation of approximately 0.06 at lag two.
%

We randomly generate 400 simulated series for each of the three processes, where the length of each series is taken to be 100. Table \ref{tab:ChiHatTable} presents the average (over the 400 simulated data sets) empirical estimator of $\chi$, from Equation (\ref{eq:ChiDef}), for each data generating process at lags $k=1,\ldots,5$. We use the empirical estimator mentioned in  \cite{Coles01a},
\begin{equation} \label{eq:chiHat}
    \hat{\chi}_k(u) =  \frac{\sum_{i=k+1}^n I\{X_i > \hat{F}^{-1}_X(u)\} I\{X_{i-k} > \hat{F}^{-1}_X(u) \}}{\sum_{i=k+1}^n I\{X_i > \hat{F}^{-1}_X(u) \}},
\end{equation}
where $I\{ \cdot \}$ represents the indicator function and $\hat{F}$ is the marginal empirical cumulative distribution function. The threshold is set at the empirical marginal 0.95 quantile (we consider other thresholds in the Supplementary Materials).
\begin{table}[ht]
\centering
\caption{The average estimate of $\chi$ (using the estimator in Equation  (\ref{eq:chiHat}) with a threshold set at their empirical 0.95 quantile) for each assumed dependent structure at lag 1 to 5.}
\label{tab:ChiHatTable}
\begin{tabular}{rrrrrr}
  \hline
 & Lag 1 & Lag 2 & Lag 3 & Lag 4 & Lag 5 \\ 
  \hline
Independent & 0.05 & 0.04 & 0.03 & 0.04 & 0.04 \\ 
GEV Process & & & & & \\
\hline
First-order & 0.30 & 0.15 & 0.08 & 0.05 & 0.04 \\ 
Markov Process & & & & & \\
\hline
MA(2) Process & 0.15 & 0.05 & 0.05 & 0.04 & 0.04 \\ 
\hline
\end{tabular}
\end{table}
We note that the independent GEV process does not allow for asymptotic dependence, and therefore we are not surprised to see the relatively small average empirical estimators at all lags  for this process. The first-order Markov GEV process does allow for asymptotic dependence, and therefore the spike at lag one is not unexpected. Although the MA(2) process does not produce true asymptotic dependence, it does induce \textit{dependence}; therefore we are not surprised to see the small spike at lag one. 

For each simulated data set, we fit two models: the stationary independent GEV model and the first-order Markov GEV model. A Bayesian inference approach is employed for each model fit, based on obtaining 2,000 posterior draws via MCMC (after burn-in). 
For each simulated data set, we estimate the true 0.95 quantile for the next realization in the series of maxima.  
In the case of the stationary independent process, this value would correspond with the 20 year return level. For the first-order Markov GEV process and MA(2) process, the interpretation is less clear; therefore, we term this parameter the \textit{conditional 0.95 quantile} (given the series of maxima). Vague Gaussian priors  are employed on the location and natural logarithm of the scale parameters. Weakly informative priors are used for the shape and dependence parameters (truncated Gaussian and Beta, respectively). For each simulated data set and for each model, the resulting posterior draws are used to generate 90\% credible intervals for the conditional 0.95 quantile. In the case of the stationary independent GEV process, we compare each 90\% credible interval with the corresponding true quantile. This true quantile of interest is based exclusively off of the known marginal GEV parameters, calculated using the GEV quantile function in Equation (\ref{eq:GEVquantile}). 
For the simulations from the first-order Markov GEV process, we also compare each credible interval with the corresponding true value, determined by the relation in Equation (\ref{eq:CondTailQtile}). In the case of the MA(2) process, (\ref{eq:MA2}) yields a simple closed form that can be used to calculate the desired true conditional 0.95 quantiles for each simulated data set. 

\begin{figure}[H]
\begin{center}
\includegraphics[width=.7\textwidth]{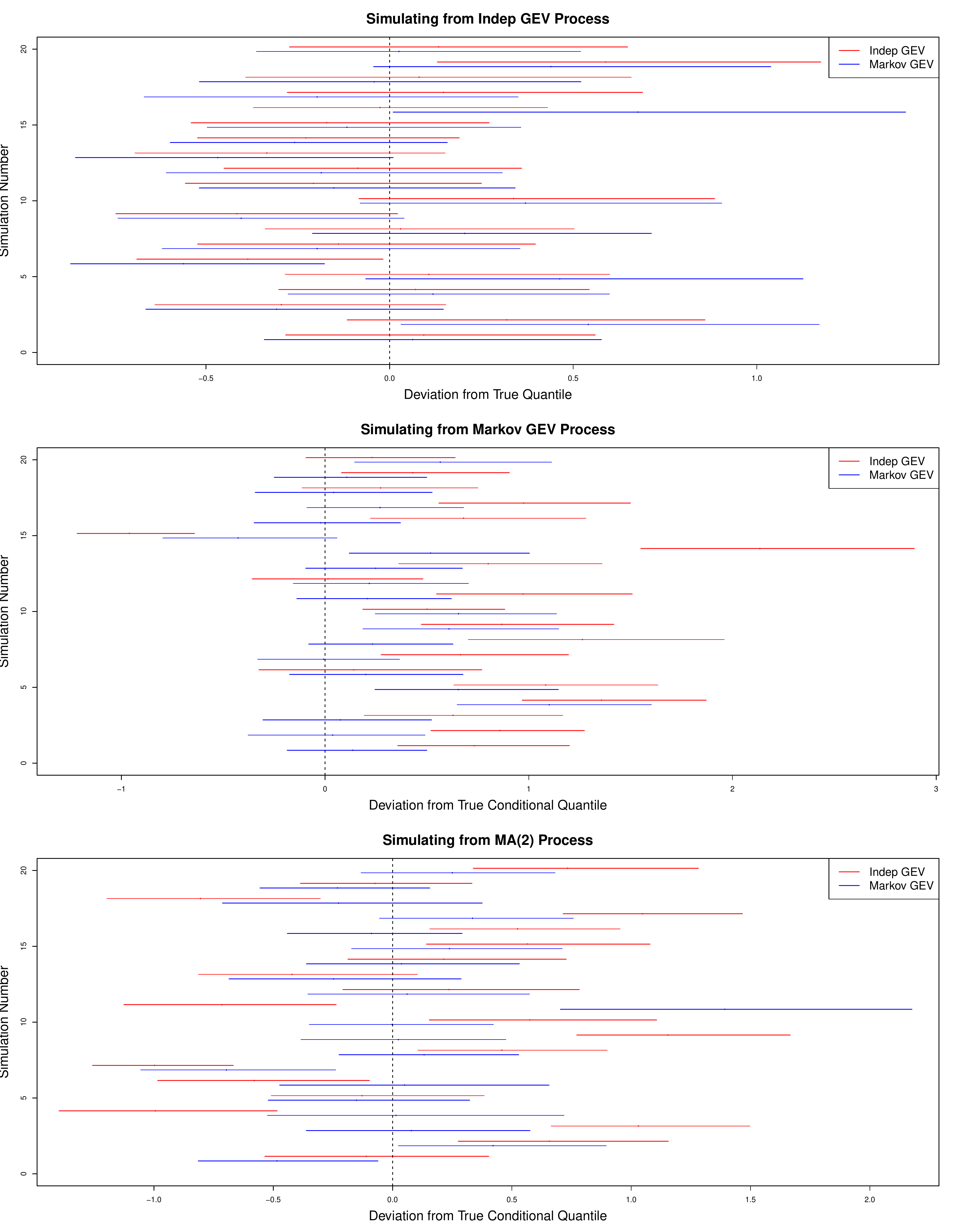}
\caption{For the first 20 simulations from each of the three process, we plot the resulting centered 90\% credible intervals for the 0.95 quantile of the next observation in the series based on both the independent GEV model and the first-order Markov GEV model. All intervals are shifted such that zero corresponds with the true 0.95 quantile value.}
\label{fig:Confint}
\end{center}
\end{figure}

The results of the simulation study are summarized in Table \ref{tab:SimStudy}. We observe that when there is no dependence present in the generating process (the stationary independent GEV process), the first-order Markov based model does not perform as well as the stationary independent GEV based model in terms of capturing the true tail quantile value. However, the dropoff in the empirical coverage rate is not large (86.5\% versus 90.8\%). 
When there is short-ranged asymptotic dependence present in the process used to generate the data (the first-order Markov GEV process), the independent GEV based model seems to perform much worse in terms of the empirical coverage rate. When sampling from the first-order Markov process, the empirical coverage rates for the 90\% credible regions are 83.7\% versus 31.4\%. Similarly, when sampling from the MA(2) process defined in Equation (\ref{eq:MA2}), the empirical coverage rates  are 83.4\% versus 42.1\%. Importantly, the first-order Markov GEV model looks to perform reasonably well despite the fact that the MA(2) process does not truly produce asymptotic dependence. For illustrative purposes, for the fist 20 simulated data sets from each of the three processes we present graphs of the resulting 90\% credible intervals based on both models in Figure \ref{fig:Confint}. Each credible interval has been centered such that its corresponding true conditional 0.95 quantile is represented by zero. 

\begin{table}[H]
\centering
\centering
\caption{The empirical coverage rate of the 90\% credible interval of the conditional 0.95 quantile for each combination of data generating process and modeling procedure.}
\label{tab:SimStudy}
\begin{tabular}{rll}
  \hline
 & First-order  &  Independent \\
 & Markov Model &  GEV Model \\ 
  \hline
Independent & 0.8651 & 0.9084 \\ 
GEV Process &  &  \\
\hline
First-order & 0.8377 & 0.3144 \\ 
Markov Process &  &  \\
\hline
MA(2) Process & 0.8337 & 0.4205 \\ 
   \hline
\end{tabular}
\end{table}

Taken together, this suggests the following set of conclusions. If a data analyst knows with certainty that the block maxima are independent, then he or she may be wise to use the independent GEV model. Similarly, if a data analyst knows with certainty that consecutive block maxima are dependent, then he or she may do better by using the first-order Markov GEV model. However, in cases when an analyst is unsure whether or not this type of dependence is present, the penalty for using the first-order Markov GEV model appears to be less onerous.

\section{Analysis of Polar Annual Minimum Air Temperature Data} \label{sect:Analysis}
Meteorological data from the Arctic and Antarctic regions of the Earth is often difficult to obtain, and may also have issues with data quality. However, these data could be interesting to analyze, as the  phenomena governing these remote regions may be different compared to more temperate regions. In this section, we present analysis of two polar annual minimum temperature time series: one in Antarctica and the other in the former Soviet Arctic region, both of which appear to exhibit short-ranged asymptotic dependence in their annual minimum near-surface temperatures. 

\subsection{Analysis of Faraday/Vernadsky station in Antarctica} \label{sect:AntarcticAnalysis}
\begin{figure}[h!]
\begin{center}
\includegraphics[width=1.0\textwidth]{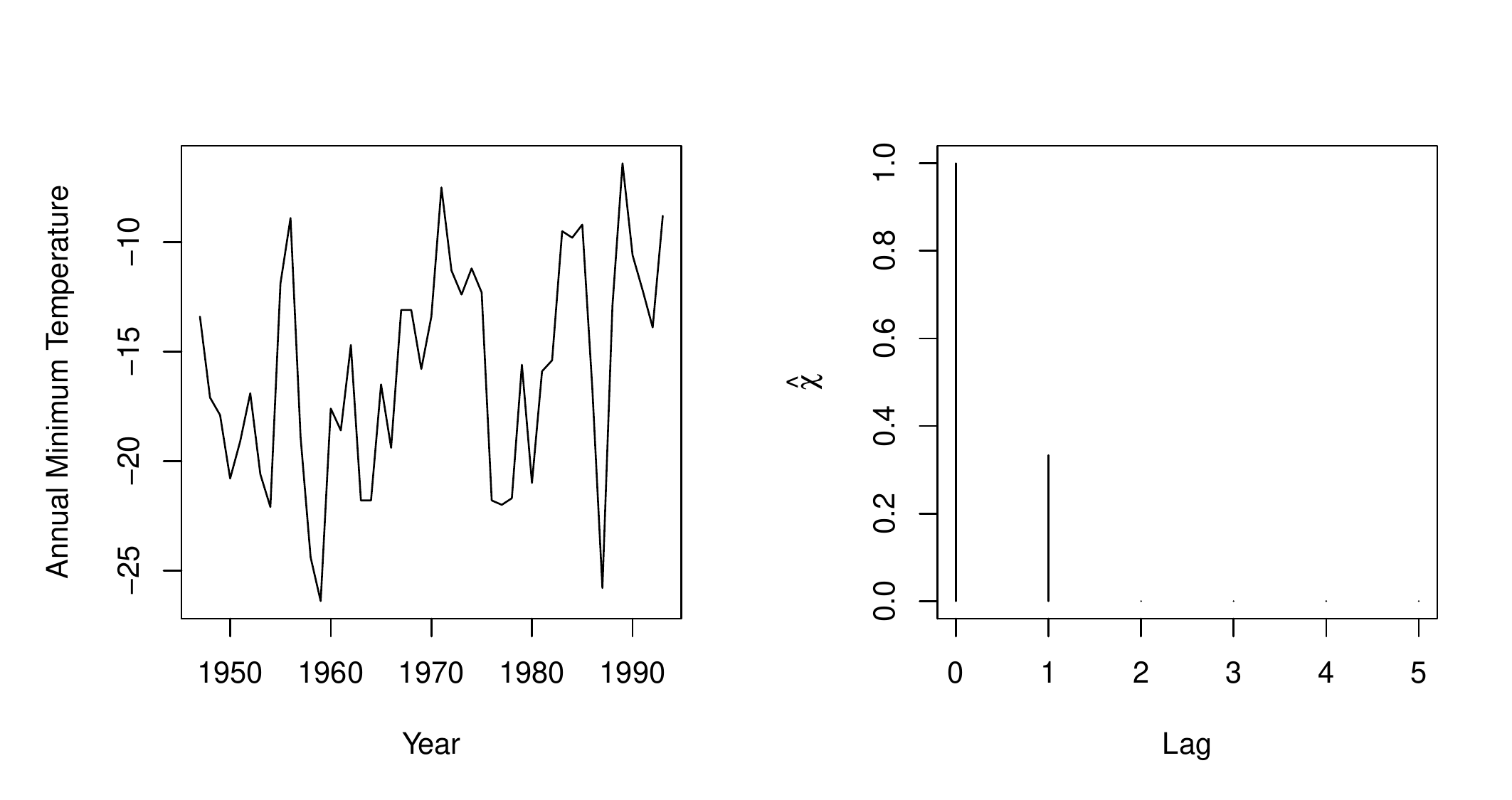}
\caption{\textbf{Left}: Annual minimum temperatures at the Faraday/Verdnansky station. \textbf{Right}: $\hat{\chi}$ at lags one through 5 for the same near-surface air temperature data, where $\chi$ is estimated using its traditional empirical estimator \citep{Coles01a} with a threshold at the empirical 0.95 quantile. 
}
\label{fig:FaradayPlot}
\end{center}
\end{figure}

We first present an analysis of an annual minimum temperature data set at the Faraday/Vernadsky station in Antarctica ($65.25^\circ$S, $64.26^\circ$W).  
This same time series is analyzed in \cite{zhu2019}, and includes data from 1947-1993 
\citep{jones2001}; we plot the corresponding  values in the left panel of Figure \ref{fig:FaradayPlot}. Based on these data, in the right panel of Figure \ref{fig:FaradayPlot} we plot the estimated value of $\chi$, using the estimator in Equation (\ref{eq:chiHat}) for $k = 1, \ldots, 5$. This plot could loosely be thought of as an asymptotic dependence analog to a sample autocorrelation plot. Here, 
the threshold is set at the empirical 0.95 quantile, but other thresholds are considered in the Supplementary Materials.  
We note that the estimated value of $\chi$ is moderately high at lag one, but drops off to zero quickly beyond this point. This exploratory analysis is not conclusive, but suggests that the first-order Markov model may be useful for these data.

\cite{zhu2019} performed analysis of this Antarctic annual minimum temperature data set, and make two overall conclusions in their work. First, they find that annual minimum temperatures seem to be increasing over this time period. Second, they conclude that dependence between annual minima exists at a lag of one year. In order to make these conclusions, they develop a time series model for (negated) annual minimum temperatures using a Gaussian dependence structure and marginal GEV distributions. To ensure that the marginal distributions are GEV, they employ probability integral transformations. The findings of \cite{zhu2019} are intriguing, and suggest an interesting climatological phenomenon; however, they are not able to determine the presence of short-ranged \textit{asymptotic} dependence. This is due to the fact that methods that incorporate Gaussian based dependence are not capable of modeling asymptotic dependence when correlation is less than one \citep{sibuya1959}. For this reason, we consider the first-order Markov GEV model developed in Section \ref{sect:DepGEV}. 

In this analysis, we use the approach developed in Section \ref{sect:DepGEV}. As in our simulation study, to allow for arbitrary GEV marginals we rely on the relation outlined in (\ref{eq:ArbitraryGEV}). In our analysis, we consider four models. In order to ensure that $\hat{\sigma} > 0$, we perform inference for the scale parameter on the log scale. To allow for temporal  non-stationarity, we consider models that include a linear temporal trend in the location parameter, i.e., $Y_t \sim \text{GEV}(\mu_0 + \mu_1 t,\sigma,\xi)$. 
This model is obtained by transforming $Z_t$, the stationary first-order Markov GEV model with unit Fr\'echet marginals, via $Y_t = \mu_0 + \mu_1 t + \sigma \left((Z_t)_+^\xi - 1\right)/\xi$ for $\mu_0, \mu_1 \in \mathbb{R}$, $\sigma > 0$, and $\xi \in \mathbb{R}$.
To this end, we consider the four models, M1 - M4, defined such that
\begin{itemize}
    \item M1: $Y_t \sim \text{GEV}(\mu,\sigma,\xi)$ - stationary independent GEV,
    \item M2: $Y_t \sim \text{GEV}(\mu_0 + \mu_1 t,\sigma,\xi)$ - independent GEV with a linear temporal trend in the location parameter,
    \item M3: $Y_t \sim \text{GEV}(\mu,\sigma,\xi)$ with dependence parameter $\alpha$ - stationary first-order Markov GEV, and
    \item M4: $Y_t \sim \text{GEV}(\mu_0 + \mu_1 t,\sigma,\xi)$ with dependence parameter $\alpha$ - first-order Markov GEV with a linear temporal trend in location parameter.
\end{itemize}
For each of the above models, a Bayesian modeling approach is taken. We use vague Gaussian priors for $\mu_0$ and $\mu_1$ in M2 and M4, and $\mu$ in M1 and M3. The scale parameter is modeled on the log scale in order to ensure positivity, and we assume a vague Gaussian prior for $\log(\sigma)$. As the GEV scale parameter is known to be difficult to estimate in practice, we employ a mildly informative truncated Gaussian prior for $\xi$ in all models. 
In M3 and M4, we use a mildly informative beta prior for $\alpha$. 
Details regarding prior distributions for all four models are available in the Appendix.

The posterior distributions do not have closed forms, thus MCMC methods are employed for inference \citep{gelman2013}.  
MCMC is performed in \verb|R| \citep{Rref} via the package \verb|rstan| \citep{rstan}, which utilizes the \verb|Stan| programming language \citep{carpenter2016}. In order to assess convergence, for each model, two independent chains with randomly selected initial values are run in parallel until 110,000 draws from each chain are obtained. The first 10,000 from each chain are discarded as burn-in and every 20$th$ observation from the remaining 100,000 draws is retained, yielding a total of 5,000 posterior draws from each chain. Convergence is assessed via traceplots (presented in Supplementary Materials) and $\hat{R}$ \citep{gelman2013}, which is approximately 1 for all parameters. Although both chains appear to have converged to the same distribution, inference hereafter is arbitrarily based  off of draws from the first chain exclusively. Additional discussion of our MCMC procedure is presented in the Supplementary Materials.

As outlined in \cite{spiegelhalter2002}, model comparison is performed by calculating the corresponding deviance information criterion (DIC) values for M1 - M4. These results indicate that M4 is the best of these four models for the Faraday/Vernadsky station data. Noting that this model includes short-ranged asymptotic dependence and a linear temporal trend in the location parameter,  
the posterior mean and key posterior quantiles for M4 are reported in Table \ref{tab:AntarcticM4postSum}.
The posterior distribution of $\alpha$ suggests that there is at least a moderate degree of asymptotic dependence between annual minima at lag one. Relying the fact that $\chi = 2 - 2^\alpha$, we transform the asymptotic dependence parameter estimate to the $\chi$ scale for the purpose of comparison. The posterior mean value for $\alpha$ is 0.657, which yields an estimated value of $\chi$ of approximately 0.423, which is quite similar the the empirical estimate at lag one presented in the right panel of Figure \ref{fig:FaradayPlot}.

We also use model output to make inference on $q_{.95}$, the annual minimum temperature associated with an exceedance probability of 0.05 for 1994. The moderate degree of dependence at lag one indicates that $q_{.95}$ may be dependent upon the annual minima in 1993, the last year of the series. Based on the MCMC output, the posterior mean estimate of the conditional tail quantile $q_{.95}$ is -18.884. For the sake of comparison, we contrast this estimate with the estimate from M2, the best fitting model that does not incorporate asymptotic dependence (based on the DIC criterion). MCMC output from M2 yields a posterior mean estimate of -21.967 for $q_{.95}$ (as seen in Table \ref{tab:ArcticM2postSum}), which is nearly 3 degrees cooler than the estimate produced by the model that does account for short-ranged asymptotic dependence.

Based on analysis of the posterior distribution of $\mu_1$, there is also moderate evidence of a linear temporal trend in the location parameter. This provides additional evidence of warming annual minimum temperatures at this location over the study period. This conclusion seems to be consistent with the findings of \cite{zhu2019} 

\begin{table}[H]%
\centering
\caption{A numerical summary of our MCMC posterior draws in M4 for the Faraday series. We report the posterior mean and several quantiles for the location intercept parameter, location temporal trend parameter, scale parameter, shape parameter, dependence parameter, and 0.95 conditional tail quantile for the next year in the series (top row to bottom row, respectively). Recall that the model is fit based on the series of \textit{negated} minima, and therefore the negative posterior mean for $\mu_1$ actually corresponds with \textit{increasing} annual minimum temperatures. 
}
\label{tab:AntarcticM4postSum}
\begin{tabular}{rrrrrrr}
  \hline
 & Mean & 2.5\% & 5\% & 50\% & 95\% & 97.5\% \\ 
  \hline
$\mu_0$ & 16.697 & 11.242 & 12.696 & 16.874 & 20.599 & 21.639 \\ 
$\mu_1$ & -0.120 & -0.280 & -0.245 & -0.123 & 0.010 & 0.051 \\ 
$\sigma$ & 5.420 & 3.510 & 3.669 & 4.737 & 7.122 & 8.371 \\ 
$\xi$ & -0.035 & -0.224 & -0.193 & -0.040 & 0.143 & 0.182 \\ 
$\alpha$ & 0.657 & 0.369 & 0.431 & 0.667 & 0.853 & 0.890 \\ 
$q_{.95}$ & -18.884 & -22.798 & -22.091 & -18.728 & -16.204 & -15.792 \\ 
   \hline
\end{tabular}
\end{table}

\begin{table}[H]
\centering
\caption{As in Table.~\ref{tab:AntarcticM4postSum} but for M2.}
\label{tab:AntarcticM2postSum}
\begin{tabular}{rrrrrrr}
  \hline
 & Mean & 2.5\% & 5\% & 50\% & 95\% & 97.5\% \\ 
  \hline
$\mu_0$ & 17.111 & 14.473 & 15.003 & 17.126 & 19.268 & 19.632 \\ 
$\mu_1$ & -0.135 & -0.226 & -0.212 & -0.136 & -0.059 & -0.043 \\ 
$\sigma$ & 4.354 & 3.496 & 3.611 & 4.319 & 5.245 & 5.415 \\ 
$\xi$ & -0.096 & -0.265 & -0.237 & -0.099 & 0.055 & 0.089 \\ 
$q_{.95}$ & -21.967 & -26.759 & -25.787 & -21.736 & -18.918 & -18.474 \\ 
   \hline
\end{tabular}
\end{table}

\begin{table}[H]
\centering
\caption{As in Table.~\ref{tab:AntarcticM4postSum} but for M4 for the Soviet series.}
\label{tab:ArcticM4postSum}
\begin{tabular}{rrrrrrr}
  \hline
 & Mean & 2.5\% & 5\% & 50\% & 95\% & 97.5\% \\ 
  \hline
$\mu_0$ & 41.169 & 39.472 & 39.784 & 41.147 & 42.596 & 42.956 \\ 
$\mu_1$ & -0.022 & -0.065 & -0.058 & -0.021 & 0.013 & 0.021 \\ 
$\sigma$ & 2.843 & 2.307 & 2.389 & 2.818 & 3.480 & 3.662 \\ 
$\xi$ & -0.141 & -0.303 & -0.282 & -0.147 & 0.016 & 0.045 \\ 
$\alpha$ & 0.813 & 0.610 & 0.644 & 0.820 & 0.958 & 0.974 \\ 
$q_{.95}$ & -45.650 & -47.792 & -47.382 & -45.573 & -44.209 & -43.978 \\ 
   \hline
\end{tabular}
\end{table}

\begin{table}[ht]
\centering
\caption{As in Table.~\ref{tab:ArcticM4postSum} but for M2.}
\label{tab:ArcticM2postSum}
\begin{tabular}{rrrrrrr}
  \hline
 & Mean & 2.5\% & 5\% & 50\% & 95\% & 97.5\% \\ 
  \hline
$\mu_0$ & 41.027 & 39.682 & 39.907 & 41.033 & 42.107 & 42.311 \\  
$\mu_1$ & -0.019 & -0.052 & -0.047 & -0.019 & 0.008 & 0.014 \\  
$\sigma$ & 2.713 & 2.284 & 2.351 & 2.707 & 3.161 & 3.258 \\  
$\xi$ & -0.170 & -0.319 & -0.295 & -0.172 & -0.035 & -0.007 \\ 
$q_{.95}$ & -46.129 & -48.278 & -47.875 & -46.048 & -44.669 & -44.433 \\ 
   \hline
\end{tabular}
\end{table}

\subsection{Analysis of Soviet Arctic Station Data}
\begin{figure}[H]
\begin{center}
\includegraphics[width=1.0\textwidth]{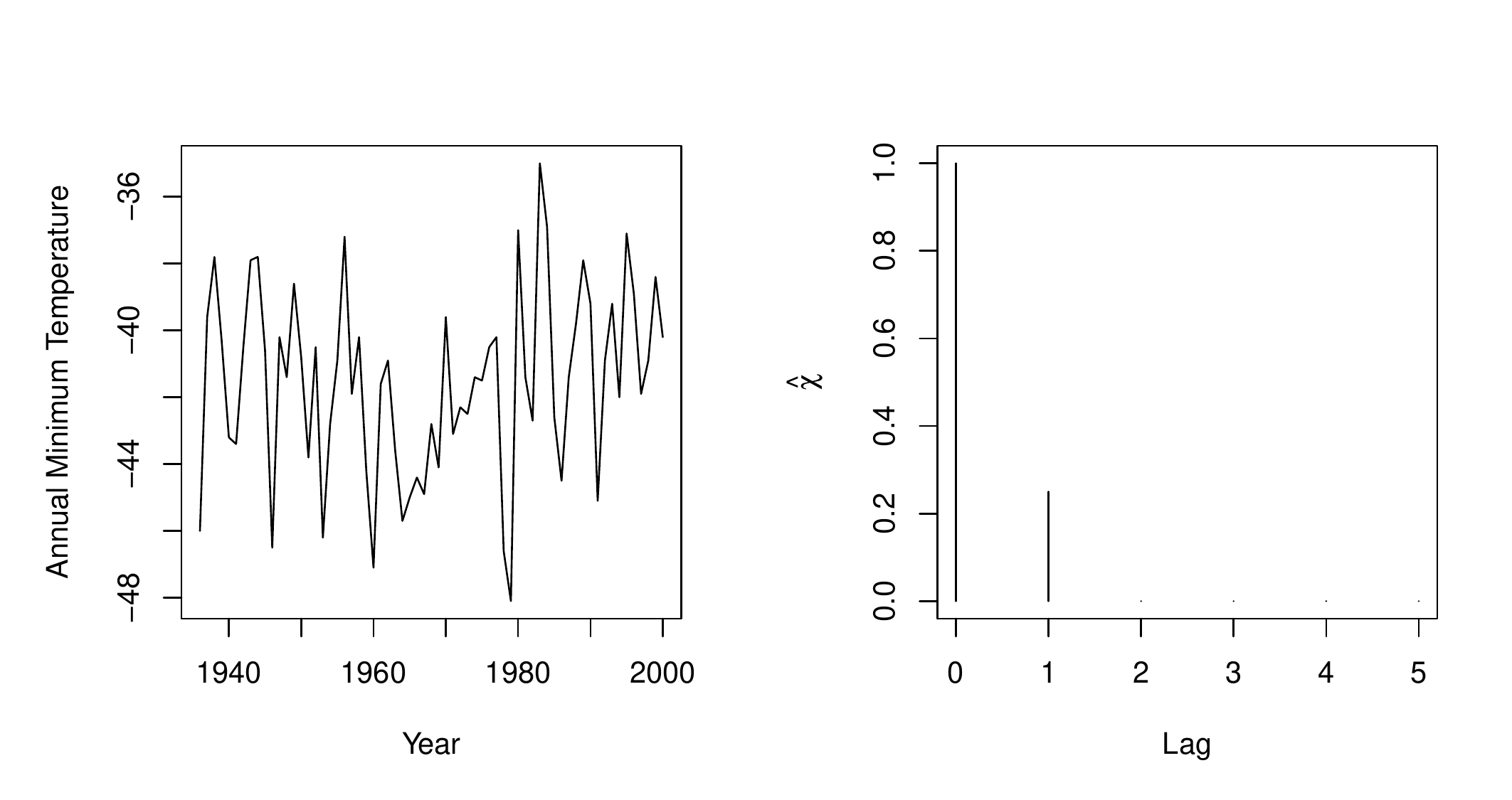}
\caption{As in Fig.~\ref{fig:FaradayPlot} but for the Soviet station.}
\label{fig:RussiaPlot}
\end{center}
\end{figure}

The analysis in Section \ref{sect:AntarcticAnalysis} indicates that there may be asymptotic dependence at lag one in annual minimum temperatures at the Faraday/Vernadsky station in Antarctica. In this section, we investigate whether this type of dependence exists at an arbitrarily selected station in the Arctic region. To this end, we perform analysis of annual minimum air temperature data using the northernmost station ($73.50^\circ$N, $80.40^\circ$E) in a Soviet research station database \citep{razuvaev1993}. We plot this time series in the left panel of Figure \ref{fig:RussiaPlot} and note that it includes data from winters (DJF) beginning in the years 1936-2000. In the right panel of Figure \ref{fig:RussiaPlot}, we plot the estimated value of $\chi$ at lags one through 5 for these data. Here, we estimate $\chi$ using the estimator described in Equation \ref{eq:chiHat} 
with the threshold at the empirical 0.95 quantile. Other thresholds are considered in the Supplementary Materials. As in the Antarctic data, the estimated value of $\chi$ is moderately high at lag one, and drops off to zero quickly beyond this point, suggesting that the first-order Markov model may be appropriate for these data as well.

In a similar fashion, we consider the Bayesian models M1 - M4 for the Arctic station data. Similar to the Antarctic data, M4 is also the best fitting model according to DIC; we find this result interesting, as it indicates short-ranged dependence may be present in these data as well. Table \ref{tab:ArcticM4postSum} presents the posterior mean and key posterior quantiles for the parameters in M4. Although the evidence appears to be less strong, these results suggest that a model that includes asymptotic dependence and a linear temporal trend in the location parameter may be favored.

Although the difference in tail quantile estimates is not as large as what we observed in the Antarctic data, we note that the estimates of $q_{.95}$  differ by nearly 1 degree for the Soviet data compared to the estimate based on M2. We note that the posterior mean and key posterior quantiles for the parameters in M2 are presented in Table \ref{tab:ArcticM2postSum}. Transforming the posterior mean asymptotic dependence parameter estimate to the $\chi$ scale results in an estimated value of $\chi$ of approximately 0.243, which is slightly lower than seen in the empirical estimate in the right panel of Figure \ref{fig:RussiaPlot}. 

\section{Discussion} \label{sect:Discussion}

In cases when a series of block maxima exhibits short-ranged temporal dependence analysis becomes more complicated. In these situations, modeling block maxima using a multivariate extreme value distribution or a max-stable process approach are valid strategies; unfortunately, these methods can be difficult to implement in practice. Max-stable must typically be fit using composite (pairwise) likelihood methods due to the fact that the joint likelihood is not tractable for even a relatively small number of observations \citep{cooley2012revstat,davison2012}. This complicates Bayesian inference approaches, though a few works have managed to implement Bayesian models in special cases \citep{reich2012,ribatet2012bayesian,thibaud2016}. Others have modeled dependence among block maxima using Gaussian copula based dependence structures \citep{zhu2019}. While straightforward to implement, a major drawback is that the Gaussian copula based approach does not allow for asymptotic dependence.


In our modeling approach, we make a first-order Markov assumption, and therefore the joint likelihood only utilizes pairwise likelihoods between consecutive observations,  which is the full likelihood of the proposed model. In situations where the dependence among block maxima is short-ranged, as we observe in our motivating data sets, this first-order Markov assumption makes a great deal of sense and results in a tractable likelihood function and utilizing a logistic structure to model dependence. 
Our model provides a simple alternative to account for short-ranged dependent block maxima.

Our motivating data sets exhibits dependence between annual minima at lag one; therefore our modeling assumptions appear to be reasonable. However, if a data set showed dependence at lags one and two, it would be possible to extend the method to assume that $X_t$ depends on $X_{t-1}$ and $X_{t-2}$. Under this scenario, the joint likelihood would then include trivariate likelihood functions. Although trivariate dependence is considerably more difficult to characterize, there are multivariate dependence structures that may make sense in these situations. We leave this extension for future work.


\section*{Conflict of Interest}
The authors declare that they have no conflicts of interest.

\section*{Acknowledgements}
Clemson University is acknowledged for its generous allotment of computing time on the Palmetto Cluster. We thank the authors of \cite{zhu2019} for sharing the Faraday/Vernadsky data set.

\section*{Data Availability Statement}
The Faraday/Verdnadsky data are described in \cite{jones2001} and have been provided to us by \cite{zhu2019}. The data used in our analysis are posted at \url{https://github.com/brooktrussell/DependentGEV/Faraday.csv}. The Soviet station data are described by 
\cite{razuvaev1993} and available at \url{https://cdiac.ess-dive.lbl.gov/ftp/ndp040/}. The data used in our analysis are posted at \url{https://github.com/brooktrussell/DependentGEV/Soviet.csv}. 

\clearpage
\appendix

\section{Additional Details of Bayesian Inference Procedure}

\begin{figure}[H]
\begin{center}
\includegraphics[width=1.0\textwidth]{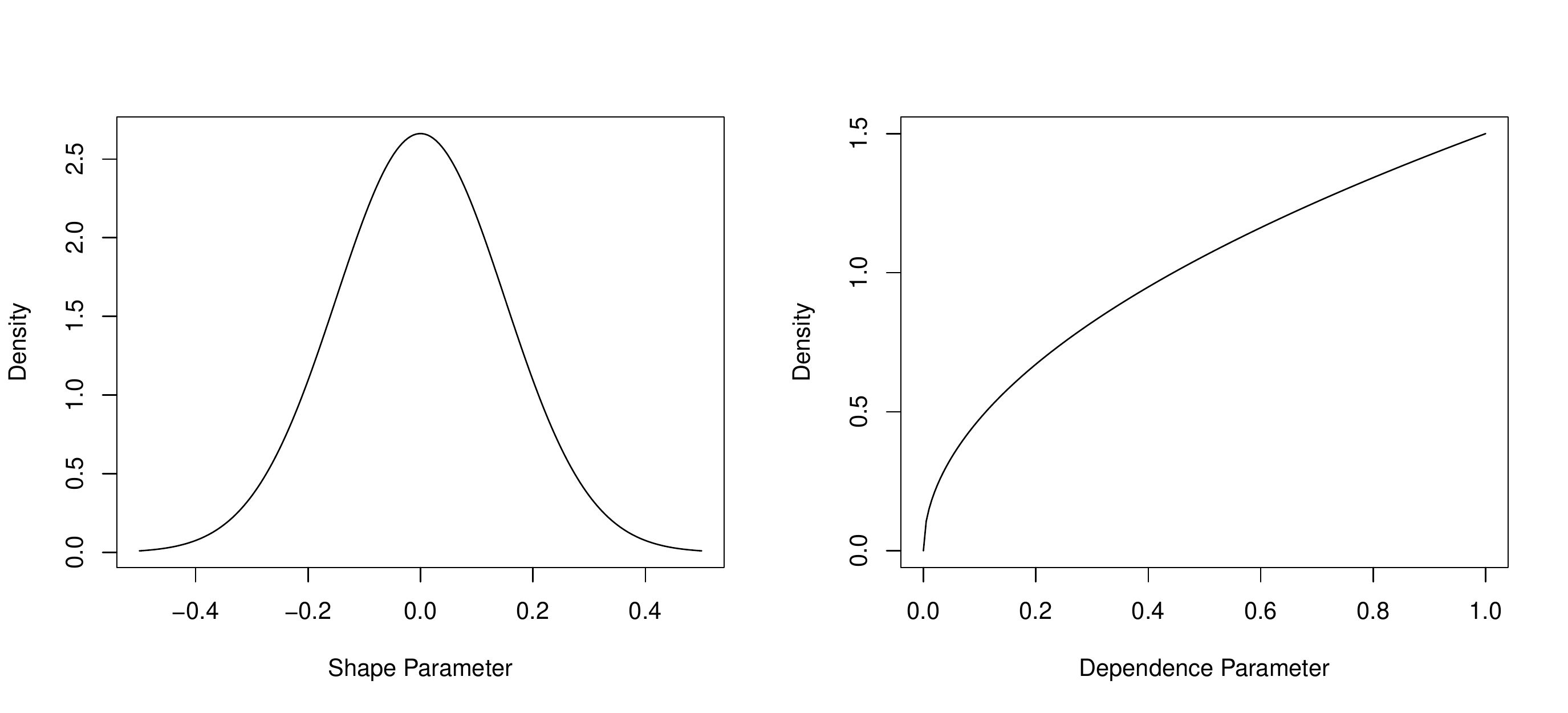}
\caption{We plot the prior distribution used for the shape parameter (L) and the dependence parameter (R).}
\label{fig:PriorPlot}
\end{center}
\end{figure}

In this section, we present additional details of our Bayesian inference procedure that were not included in the manuscript. In our Bayesian modeling procedure, we utilize the following prior distributions for model parameters in M1-M4.
\begin{itemize}
    \item M1: $\mu \sim N(0,100^2)$, $\log(\sigma) \sim N(0,15^2)$, and $\xi \sim TN(0,0.15^2,-0.5,0.5)$
    \item M2: $\mu_0 \sim N(0,100^2)$, $\mu_1 \sim N(0,15^2)$, $\log(\sigma) \sim N(0,15^2)$, and $\xi \sim TN(0,0.15^2,-0.5,0.5)$
    \item M3: $\mu \sim N(0,100^2)$, $\log(\sigma) \sim N(0,15^2)$, $\xi \sim TN(0,0.15^2,-0.5,0.5)$, and $\alpha \sim Beta(1.5,1)$
    \item M4: $\mu_0 \sim N(0,100^2)$, $\mu_1 \sim N(0,15^2)$, $\log(\sigma) \sim N(0,15^2)$, $\xi \sim TN(0,0.15^2,-0.5,0.5)$, and $\alpha \sim Beta(1.5,1)$
\end{itemize}
We note that $TN$ denotes the truncated Gaussian distribution where the parameters are the mean, variance, lower truncation value, and upper truncation value (respectively). The prior distribution for the shape parameter $\xi$ and the dependence parameter $\alpha$ are plotted in the left and right (respectively) panels Figure \ref{fig:PriorPlot}.

In order to minimize the effect of temporal dependence in MCMC realizations, we thin by retaining every 20$th$ observation. This strategy appears to be reasonably effective, as the number of effective observations for each parameter (as calculated by the \verb|rstan| package) are reasonably large, as seen in Table \ref{tab:NumEff}. 

\begin{figure}[H]
\begin{center}
\includegraphics[width=1.0\textwidth]{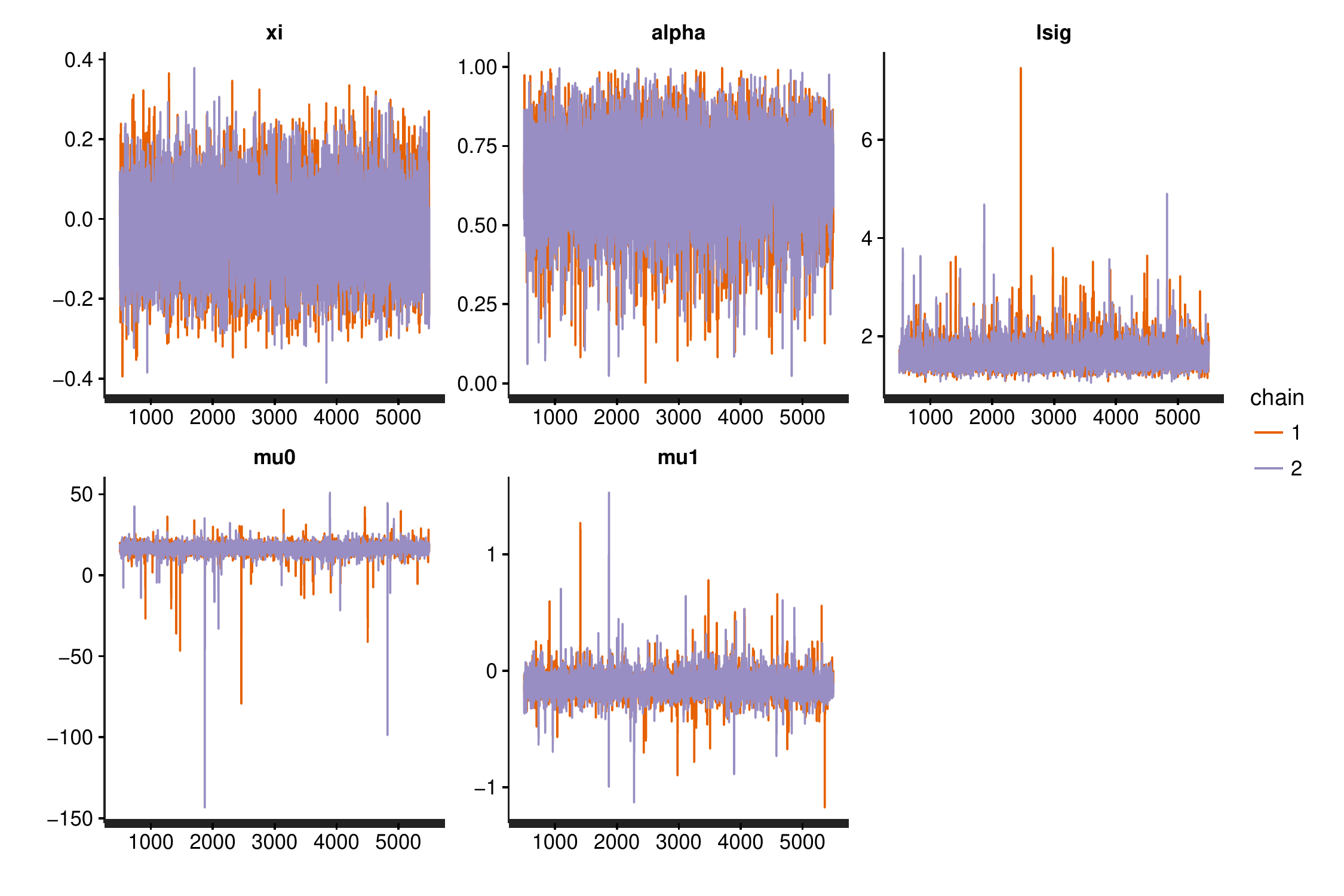}
\caption{In order to aid in assessing convergence, we present traceplots for the parameters in M4 based on the Faraday station data.}
\label{fig:FaradayTraceplots}
\end{center}
\end{figure}

\begin{figure}[H]
\begin{center}
\includegraphics[width=1.0\textwidth]{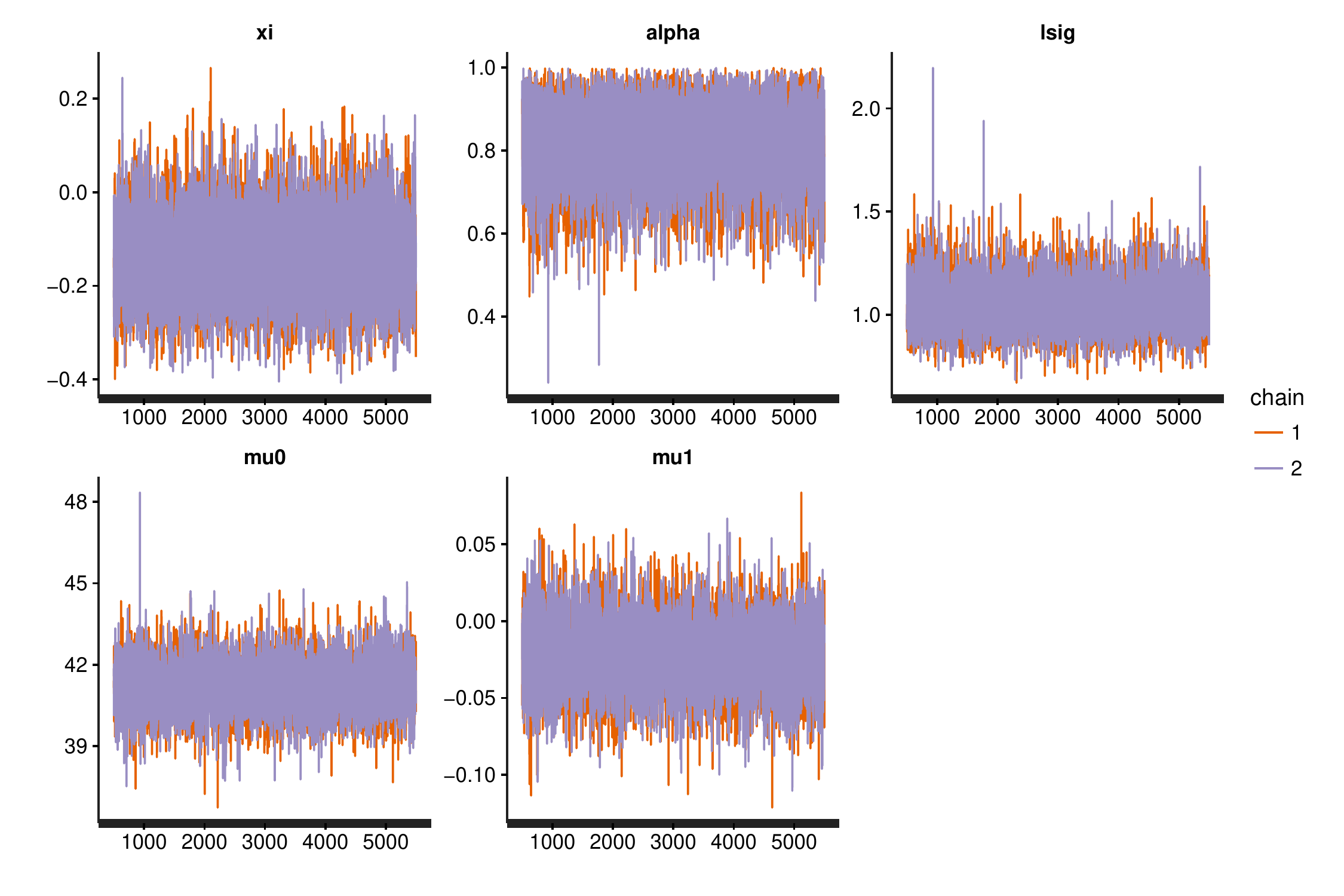}
\caption{In order to aid in assessing convergence, we present traceplots for the parameters in M4 based on the Soviet station data.}
\label{fig:SovietTraceplots}
\end{center}
\end{figure}

We also believe that it is reasonable to think that both chains have converged. In the manuscript, we mention that the value of $\hat{R}$ is approximately one for all parameters. This is further evidenced by the traceplots in Figures \ref{fig:FaradayTraceplots} and \ref{fig:SovietTraceplots}, which appear to be consistent with convergence. 

\begin{table}[ht]
\centering
\caption{We report the number of effective MCMC realizations, recalling that a total of 10,000 draws were retained after thinning and burn-in (including both chains).}
\label{tab:NumEff}
\begin{tabular}{rrrrrr}
  \hline
 & $\xi$ & $\alpha$ & $\sigma$ & $\mu_0$ & $\mu_1$ \\ 
  \hline
Soviet Station & 10,000 & 9,643 & 9,360 & 10,000 & 10,000 \\ 
\hline
Faraday Station & 9764 & 9169 & 8526 & 8701 & 9427 \\ 
\hline
\end{tabular}
\end{table}

As described in the manuscript's Appendix, we employ vague Gaussian priors for the location and log scale parameters. The GEV shape parameter is known to be difficult to estimate; therefore, we use a slightly informative truncated Normal prior distribution for this parameter. We believe that a prior that places no mass outside of $(-0.5,0.5)$ is reasonable for this application. Since we believe that the dependence parameter $\alpha$ is moderate, we employ a Beta prior that has much higher density over $(0.5,1.0)$ compared to the interval $(0,0.5)$. For the sake of comparison, the prior densities for $\alpha$ and $\xi$ are plotted in Figure \ref{fig:PriorPlot}.


\section{A Comparison of Different Thresholds}
To supplement the results in the the manuscript, we consider different threshold values. Table \ref{tab:ChiHatTable925} is analogous to Table 1 in the manuscript, except that it is based on thresholding at the empirical 0.925 quantile. Table \ref{tab:ChiHatTable90} is analogous to Table 1 in the manuscript, except that it is based on thresholding at the empirical 0.90 quantile. Both tables convey similar results compared to Table 1 in the manuscript.

\begin{table}[ht]
\centering
\caption{The average estimate of $\chi$ (using the estimator in Equation  (20) with a threshold set at their empirical 0.925 quantile) for each assumed dependent structure at lag 1 to 5.}
\label{tab:ChiHatTable925}
\begin{tabular}{rrrrrr}
  \hline
 & Lag 1 & Lag 2 & Lag 3 & Lag 4 & Lag 5 \\ 
  \hline
Independent & 0.07 & 0.07 & 0.07 & 0.08 & 0.08 \\ 
GEV Process & & & & & \\
\hline
First-order & 0.36 & 0.19 & 0.12 & 0.10 & 0.08 \\ 
Markov Process & & & & & \\
\hline
MA(2) Process & 0.37 & 0.14 & 0.08 & 0.07 & 0.07 \\
\hline
\end{tabular}
\end{table}

\begin{table}[ht]
\centering
\caption{The average estimate of $\chi$ (using the estimator in Equation  (20) with a threshold set at their empirical 0.90 quantile) for each assumed dependent structure at lag 1 to 5.}
\label{tab:ChiHatTable90}
\begin{tabular}{rrrrrr}
  \hline
 & Lag 1 & Lag 2 & Lag 3 & Lag 4 & Lag 5 \\ 
  \hline
Independent & 0.09 & 0.09 & 0.09 & 0.09 & 0.10 \\ 
GEV Process & & & & & \\
\hline
First-order & 0.38 & 0.22 & 0.15 & 0.12 & 0.11 \\ 
Markov Process & & & & & \\
\hline
MA(2) Process & 0.39 & 0.16 & 0.10 & 0.10 & 0.09 \\
\hline
\end{tabular}
\end{table}

\begin{figure}[H]
\begin{center}
\includegraphics[width=1.0\textwidth]{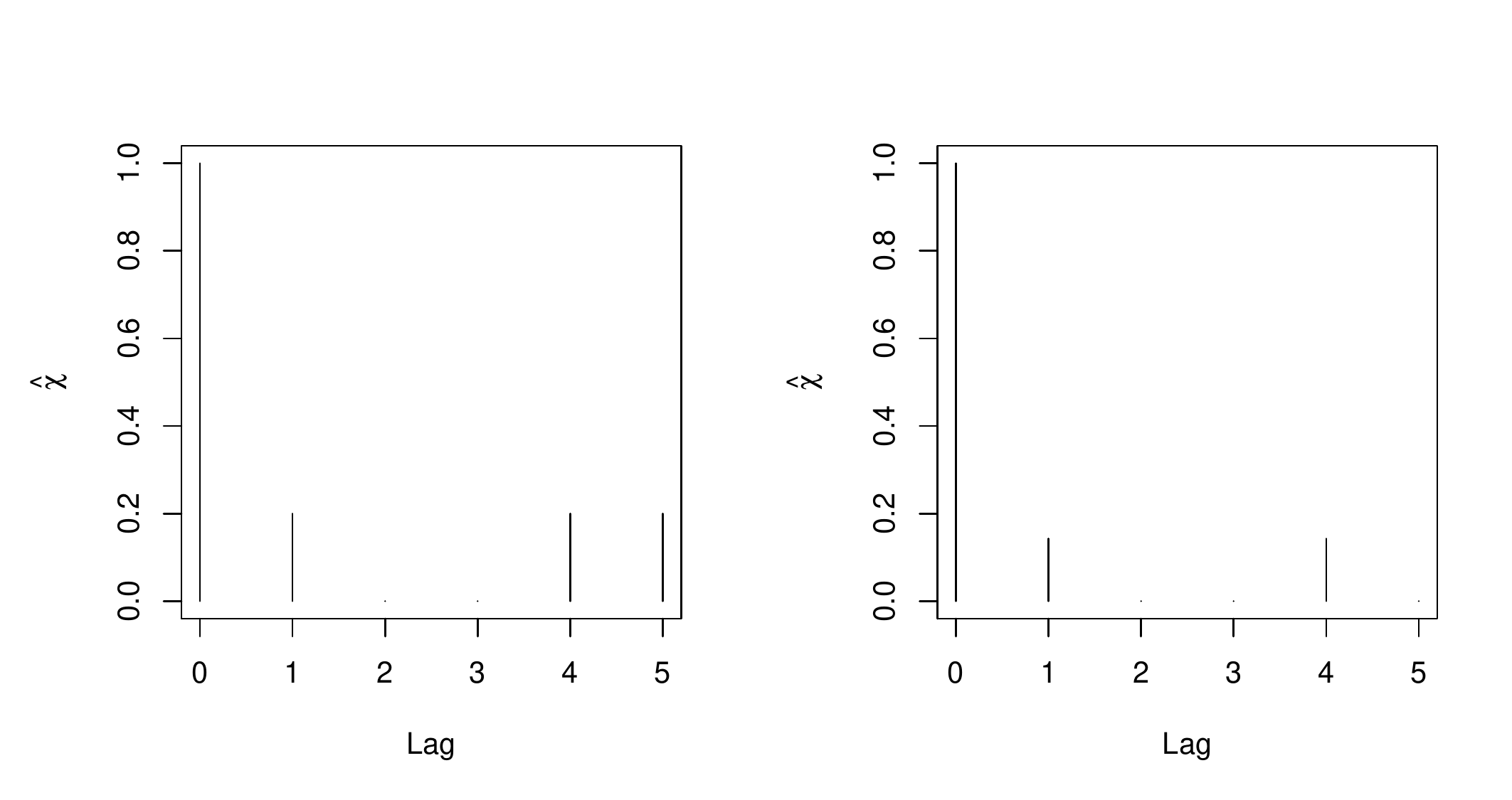}
\caption{We plot the empirical estimates of $\chi$ at lags one through five, based on the empirical 0.90 quantile, for the Faraday data (L) and the Soviet data (R).}
\label{fig:90qtile}
\end{center}
\end{figure}

\begin{figure}[H]
\begin{center}
\includegraphics[width=1.0\textwidth]{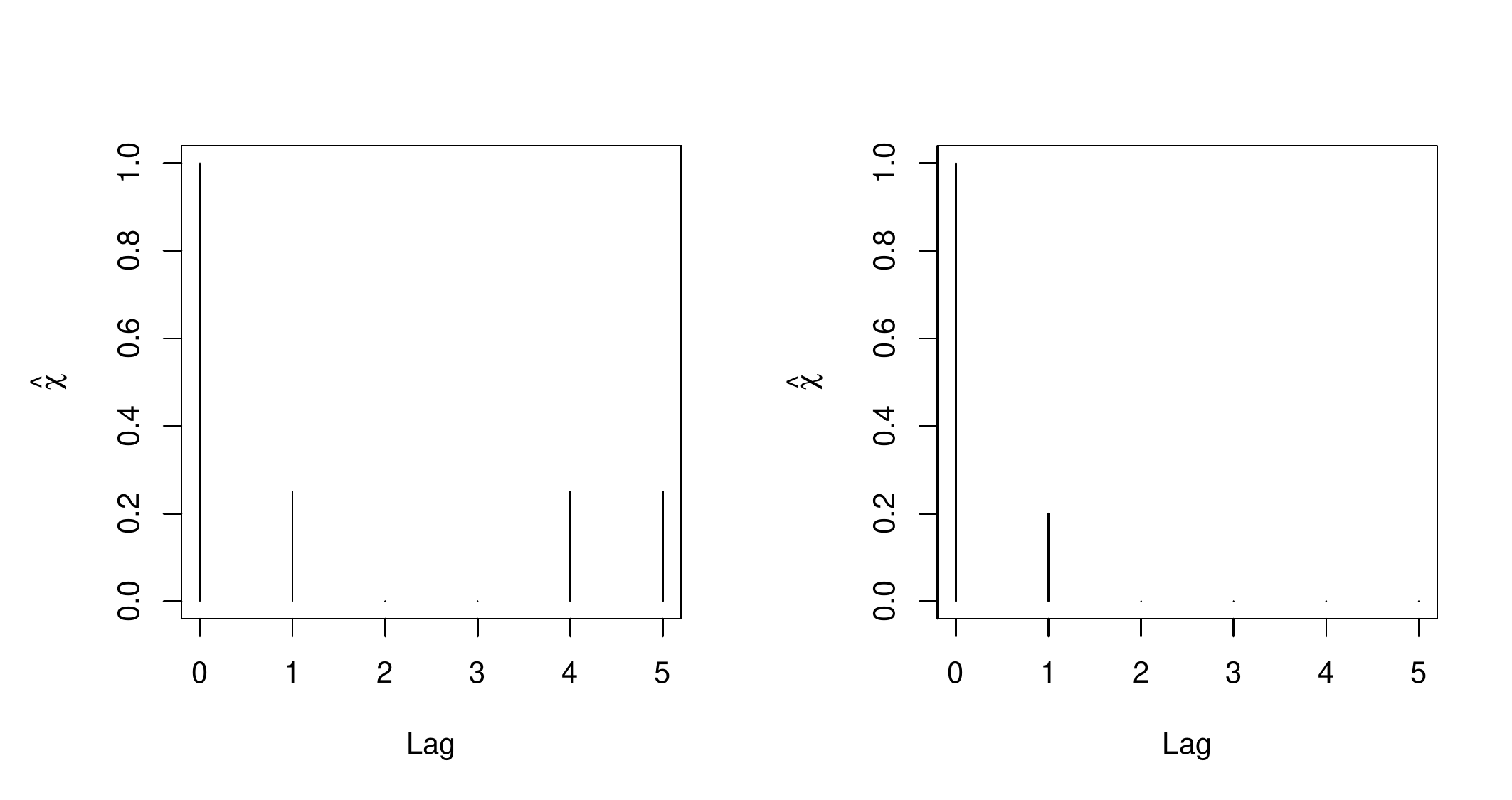}
\caption{We plot the empirical estimates of $\chi$ at lags one through five, based on the empirical 0.925 quantile, for the Faraday data (L) and the Soviet data (R).}
\label{fig:925qtile}
\end{center}
\end{figure}

In order to investigate the degree to which the graphs in the right panels of Figures 2 and 3 in the manuscript are threshold dependent, we create additional figures based on alternative thresholds. The graph in the left panel of Figure \ref{fig:90qtile} is analogous to the right panel of Figure 2 in the manuscript, but thresholds at the empirical 0.90 quantile. The graph in the right panel of Figure \ref{fig:90qtile} is analogous to the right panel of Figure 3 in the manuscript, but thresholds at the empirical 0.90 quantile. The graph in the left panel of Figure \ref{fig:925qtile}is analogous to the right panel of Figure 2 in the manuscript, but thresholds at the empirical 0.925 quantile. The graph in the right panel of Figure \ref{fig:925qtile} is analogous to the right panel of Figure 3 in the manuscript, but thresholds at the empirical 0.925 quantile. All graphs presented here seem to yield similar conclusions as presented in the manuscript.

\bibliographystyle{apalike}
\bibliography{myrefs}

\end{document}